# Observation of high efficiency Betatron radiation from femtosecond petawatt laser irradiated near critical plasmas

J. H. Tan[1, 2], Y. F. Li[1, a)], D. Z. Li[4], J. Feng[2], Y. J. Li[2], W. J. Zhou[2], Y. H. Yan[6], Z. M. Zhang[6], B. L. Zhang[1, 5], J. G. Wang[1], Y. Q. Gu[6], Y. T. Li[1, 3, 5, 7], L. M. Chen[2, 3, b)]

[1] Beijing National laboratory for Condensed Matter Physics, Institute of Physics, Chinese Academy of Sciences, Beijing 100190, P. R. China
[2] School of Physics and Astronomy, Shanghai Jiao Tong University, Shanghai 200240, P. R. China
[3] IFSA Collaborative Innovation Center, Shanghai Jiao Tong University, Shanghai 200240, P. R. China
[4] Institute of High Energy Physics, Chinese Academy of Sciences, Beijing 100049, P. R. China
[5] School of Physical Sciences, University of Chinese Academy of Sciences, Beijing 100049, P. R. China
[6] Laser Fusion Research Center, China Academy of Engineering Physics, Mianyang 621900, P. R. China
[7] Songshan Lake Materials Laboratory, Dongguan, Guangdong 523808, P. R China

a) yflx@iphy.ac.cn, b) lmchen@sjtu.edu.cn

We present an experimental demonstration of high conversion efficiency Betatron x-ray radiation from petawatt laser irradiated near critical plasmas. Direct laser acceleration serves as the dominant regime when laser pulse of ~5×10$^{20}$ W/cm$^2$ intensity is focused into plasmas with electron density of 3×10$^{20}$ /cm$^3$. Electron beam with a charge of ~35 nC is accelerated up to a maximum energy of 70 MeV and emit x-rays when oscillating in the laser field. The deduced energy conversion efficiency from laser to x-rays is up to 10$^{-4}$, orders of magnitude higher than other betatron regimes. Enhancement of acceleration and radiation with sharp plasma density boundary is also obtained and further interpreted with 2D particle-in-cell simulations.

Probing ultra-fast atomic scale transition of materials under ultra-short pump relies mainly on x-rays with femtosecond duration and high-brightness, which can be routinely provided by national scale synchrotron radiation and x-ray free electron lasers facilities. Table-top betatron x-ray pulses generated from laser plasma acceleration also offer great potentialities because of its ultra-short time duration, micron source size and high brightness [1], and is recently proved to facilitate time resolved x-ray absorption spectroscopy (XAS) diagnostics of the nonequilibrium dynamics of metal samples at extreme temperature and pressure (Warm Dense Matter) [2]. A rise-time of electron temperature below 100 fs is resolved before the thermalization of crystal lattice. In the field of x-ray microcomputed tomography (μCT), betatron radiation is also utilized to circumvent the trade-off between spatial resolution and acquisition time [3,4].

Despite above remarkable features, studies of the two typical field to date is still far from the ideal. In the context of XAS experiments, accumulation of tens of shots per absorption spectrum is required for a better signal to noise ratio. This stems from two considerations that the single shot number of x-ray photons in the absorption band is not enough and the intrinsic spectral flux fluctuation of betatron radiation is relatively large [2]. The same requirement also exists in μCT studies, where multi-shots are acquired for each sample rotation angle [3]. One intuitive solution to above problems is to acquire absorption spectrum or μCT image in a single shot scenario. It is, therefore, necessary to explore new regime for betatron radiation that generating 2-3 orders of magnitude more x-ray photons can be possible.

Betatron radiation originates from deflecting of electron beam by the transverse quasistatic electromagnetic field or/and the laser field when accelerated in the plasma channel [1,5], The total photon number $N_t = N_\gamma N_e N_\beta$ scales with the yield per electron per betatron period $N_\gamma = 5\pi\alpha K_\beta/3^{1/2}$, electron number $N_e$ and average number of periods $N_\beta$, where $\alpha = 1/137$ is the fine structure constant and $K_\beta = \gamma_e \omega_\beta r_\beta/c$ is the strength parameter indicating the normalized transverse momentum of electron with $\omega_\beta \sim \omega_p/(2\gamma_e)^{1/2}$ the betatron frequency, $\omega_p$, $\gamma_e$, $r_\beta$ the plasma frequency, electron Lorentz factor and betatron motion amplitude, respectively. Direct laser acceleration (DLA) regime dominates when laser pulses with tens of fs duration and petawatt peak power [6,7], corresponding to a peak normalized vector potential $a_0 = 0.89 I_l^{1/2}[10^{18} W/cm^2]\lambda_l[\mu m]$ about 10-25, incident into near critical density plasma (NCD), $n_e \sim n_c$, where $I_l$ is the laser intensity and $n_e, n_c = \epsilon_0 m_e \omega_l^2/e^2$ are the plasma density and critical density of laser wavelength $\lambda_l = 2\pi c/\omega_l$. In DLA, Electrons directly gain energy from laser electric field when betatron frequency coincide with the doppler shifted laser frequency $\omega_l$ [6], i.e., $\omega_\beta = \omega_l(1 - v_x/v_{ph})$ where $v_x$ is the electron velocity along laser propagation axis and $v_{ph}$ is the laser phase velocity. One can immediately note that the resonant condition can be easier fulfilled at high plasma density. Consequently, enormous electrons could be accelerated to high energies and strongly wiggled directly by laser field when fs PW laser pulses interacting with NCD plasmas, thus generating abundant x-ray photons in a single shot.

In previous studies, betatron radiation from DLA by sub-ps laser and 10$^{19}$ cm$^{-3}$ plasma is observed to scale as $a_0$ and the interaction length [8]. Recent simulations predict that up to 1% laser energy can be converted to MeV gamma-rays by fs laser of 10$^{21}$ W/cm$^2$ intensity interacting with NCD plasmas [9]. Further, at intensities beyond 10$^{22}$ W/cm$^2$, where quantum electrodynamic effects appear to be significant, the conversion efficiency can be enhanced by a factor of 5-20.

In this Letter, we report the first observation of betatron radiation from femtosecond petawatt (PW) laser irradiated gaseous near critical plasmas. We show that the gas density gradient affects the electron acceleration and photon yield by determining the channel formation thus acceleration distance through experimental measurements and 2D particle-in-cell (PIC) simulations. For sharp density boundary, this PW and NCD scenario increases x-ray photon yield by orders of magnitude with respect to self-modulated or laser wake-field acceleration. Thus, this may pave the way to overcome above mentioned limitations, with the possibility of exploiting single shot measurements of XAS spectrum or μCT image.

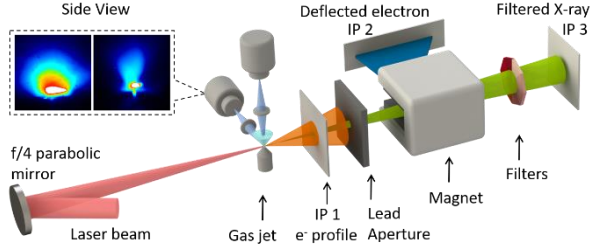
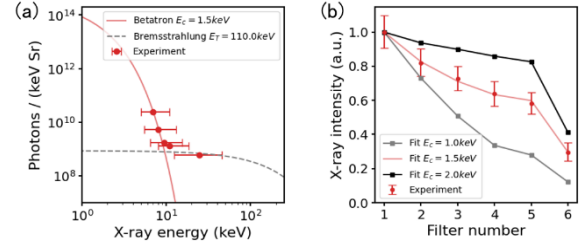

FIG. 1. Schematic of the experiment setup. The laser pulse is focused by a f/4 on-axis parabolic mirror upon a specially designed gas jet, producing electron (blue) and x-ray (green) beams. The angular distribution of electron beam is recorded on image plate 1 (IP 1) along the laser propagation axis (28 cm from the source). There is also an aperture on IP 1 for the spectrum measurements of electron and x-ray beams. Electron beam (blue) passing through a lead aperture is then dispersed by a 0.9 T dipole magnet and finally recorded on IP 2. X-rays (green) passing through the lead aperture is then filtered by an 8 metal filters onto IP 3 (51.5 cm from the source). The interaction point is view by two Questar telescope system (top and side view) and the up left inset shows the raw images of the side-viewed gas plume with poor (left) and best (right) gas confinement.

FIG. 2. (a) Measured (red dot) and deduced (red solid line, $E_c = 1.5\ keV$, $E_b = 110\ keV$, $A = 3.35 \times 10^{-6}$) betatron and bremsstrahlung (dashed gray line) spectrum with a set of cutoff filters for $a_0 = 13.5$, $n_e = 5.2 \times 10^{20}\ cm^{-3}$ and $L_{grad} \sim 100\ \mu m$ shock nozzle. (b) Measured (red dot with error bar) and calculated x-ray intensity through different filters on IP 3. The gray, red and black dots represent $E_c=1.0$ keV, 1.5 keV and 2.0 keV, respectively. The filters, numbered 1-6, are 40 μm Al, 55 μm Al, 85 μm Al, 160 μm Al, 240 μm Al and 200 μm Cu with 40 μm Al.

The experiments were conducted using SILEX-II laser facility [10] at Laser Fusion Research Center, China Academy of Engineering Physics. SILEX-II delivers pulses of 30±10 J energy, 30±5 fs duration after compression and is then focused by a f/1.7 on-axis parabolic mirror which resulting in a 3.8 μm waist gaussian focal spot. The gas target is produced by symmetric shock nozzles, which form a dense and narrow gas profile at certain height by generate converging oblique shocks (see reference [11] for details), pumped by a SL-GT-10 high density system from SourceLAB. The gas density was calibrated off-line with interferometer and can be well described by combination of a gaussian and an inverse quadratic function with a characteristic length $L_{grad}$ indicating the length at which the density decreases to about 1/10 of the maximum. In our experiment, pure nitrogen was used and the peak atomic density can be as high as $1\times10^{21}$ cm$^{-3}$. Two Questar telescope systems were used to monitor the scattering and/or self-emission light in side and top directions with a spatial resolution about 7 μm. The angular distribution of electron beam was recorded by an image plate (IP 1) which has a hole in the center for subsequent electron spectrum measurement. The electron beam, then passing through a lead aperture, is deflected onto image plate 2 (IP 2) by a dipole magnet (0.9 T, 8 cm). The magnetic field distribution and geometrical layout is considered for better accuracy of spectrum calculation. To characterize the x-rays, we have used two sets of metal filters, one consists of six filters with 1/e cutoff energy from keV to about 30 keV, another is arranged as Ross pairs [12] for diagnosing of x-rays up to 100 keV. Note that only one set of filters is used in a single shot.

The spectral characteristics of betatron radiation can be described by wiggler type synchrotron spectrum when the strength parameter satisfies $K_\beta \gg 1$. This suits the experimental conditions well especially while electrons directly interact with ultra-relativistic laser field. Despite the wiggler spectrum, similar to Ref. [13], we assume an additional bremsstrahlung background, so that the on axis differential spectrum is:

$$\frac{dN}{dE} \propto \frac{1}{E}\left(\frac{E}{E_c}\right)^2 K_{2/3}^2\left[\frac{E}{E_c}\right] + A\exp\left[-\frac{E}{E_b}\right]$$

where $E_c \simeq 3\hbar K_\beta \gamma_e^2 \omega_\beta$ is the critical energy that the total energy radiated below and above are equal, $E_b$ is the effective temperature of bremsstrahlung spectrum, A is the amplitude ratio of betatron radiation and bremsstrahlung, and $K_{2/3}$ is modified Bessel function of the second kind.

Then we take the filter transmission $T_i(E)$ and the calibrated image plate response [14] $R(E)$ into account and integrate to yield the theoretical photo-stimulated luminescence (PSL) of each filter as $PSL_i^t = \int (dN/dE)\, T_i(E) R(E) dE$. Lastly, reconstruction of the betatron spectrum turns out be an optimization problem subject to the experimental $PSL_i^e$, and we adopt a least mean square fit to minimize the residual $\sum_i (PSL_i^t/PSL_1^t - PSL_i^e/PSL_1^e)^2$. FIG. 2(a) shows an example best fit spectrum for $n_e = 5.2 \times 10^{20} cm^3$ with $E_c = 1.5\ keV$, $E_b = 110\ keV$ and $A = 3.35 \times 10^{-6}$, which is in good agree with the experimental data. Comparison of the calculated and measured x-ray intensity, under variation of critical energy, also justifies the treatment of spectrum. The theoretical x-ray intensity for each filter with $E_c = 1.0\ keV, 1.5\ keV$ and $2.0\ keV$ is shown in FIG. 2(b). The spectrum is measured with the low energy cutoff filters while the Ross filters are effective in energy band where bremsstrahlung dominates ($E > 10\ keV$). From the sharp filter edges, we deduced an x-ray source size about 200 μm. Note that it is a very upper limit because IP 3 located immediately behind the filter and poor magnification would be obtained. However, this indicates that the majority x-rays come from the gas jet.

One can further calibrate the absolute amplitude of spectrum to obtain the total yield. Under optimum condition, the total photon number is found to be more than $2.1\times10^{13}$ (>1 keV) and $2.3 \times 10^9$ (>10 keV) considering the same divergence angle with the electron beam θ ~ 35° (~ 0.3 Sr). The total energy of x-ray is above 6 mJ (>1 keV), which corresponding to an energy conversion efficiency of $3 \times 10^{-4}$. This is, to the best of our knowledge, the highest efficiency of betatron radiation source achieved to date.

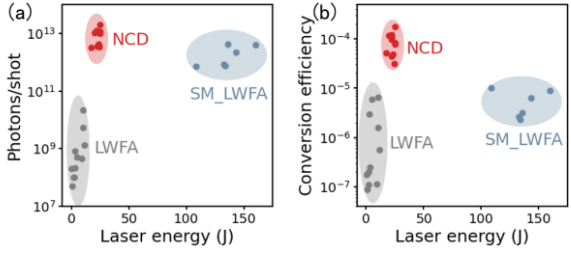

FIG. 3. single shot (a) x-ray photon yield (b) laser to x-ray energy conversion efficiency of betatron radiation generated from LWFA, SM-LWFA and NCD regimes. Data of LWFA and SM-LWFA are adapted from 13 experiments (see text for details).

In contrast to laser wakefield acceleration (LWFA) [3,4,15-23] and self-modulated laser wakefield acceleration (SM-LWFA) [13,24], our scheme, fs petawatt laser irradiated NCD plasmas, enhances the betatron radiation in total photon yield as well as laser to x-ray energy conversion efficiency (FIG. 3) This is the consequence of two effects. First, in DLA scenario, laser pulse overrides the resonant electrons one laser wavelength in each betatron period. While NCD helps to slow down the laser, the requirement for electron velocity $v_x$ to catch up with the laser pulse will be relaxed and bulk electrons will hit the resonance and be accelerated [25]. The total charge of electrons associated with FIG.2(a) is estimated to be 36.6 nC. This is about 1000 times higher than the tens of pC charge of LWFA and comparable to that of SM-LWFA. Second, the total x-ray number $N_t$ and energy $N_t E_c$ increase linearly and quadratically with $r_\beta$. As a consequence of betatron resonance with PW laser field, exchange of transverse momentum may lead to a considerable increase of $r_\beta$. For electrons beam with Maxwellian spectrum, we estimate the average betatron oscillation amplitude by $\langle r_\beta \rangle \simeq E_c c / 3\hbar \langle \gamma_e \rangle^3 \omega_\beta^2$, where $\langle \gamma_e \rangle = T_e$ is the effective electron temperature. For electron beam with $7.0 < \gamma_e < 9.8$ (see text below), $\langle r_\beta \rangle \simeq 4.2 \pm 1.4$ μm is estimated. This estimation is approaching the radius of plasma channel, which is the upper limit for $r_\beta$ [7,26], observed in particle in cell simulation.

In addition, it is also noted that the energy conversion efficient can be further enhanced about 2 orders of magnitude with MeV critical energy according to simulations [9]. This is attributed to the lower electron energy in our experiment since the total energy radiated scales with electron energy as $\sim N_t E_c \propto \gamma_e^4 \omega_\beta^3 r_\beta^2$. Electrons follow a Maxwellian spectrum distribution with several MeV effective temperature and extend to tens of MeV. However, gas density gradient is found to be an important factor on electron energy, and is investigated by varying the characteristic length of shock nozzles, i.e., $L_{grad}$ ~ 100 μm, 200 μm for Case I, II, and a cylindrical free expanding nozzle of 400 μm diameter, Case III.

FIG. 4(a) shows example spectrum of above three cases under approximately same laser conditions. Electron temperature is observed to increase with sharper density gradient, i.e., increasing from $T_e = 1.9\ MeV$ and a cutoff energy about 20 MeV for Case

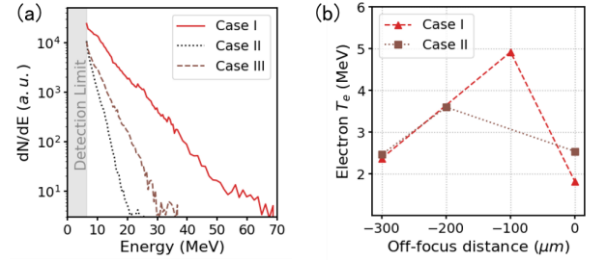

FIG. 4. (a) Electron energy spectrum measured under three gas density gradient conditions for $a_0 = 14.9 \pm 1.5$ and $n_e \sim 5 \times 10^{20} cm^{-3}$, where Case I, II, III represent shock nozzles with characteristic length $L_{grad}$ ~ 100 μm, 200 μm and a cylindrical nozzle of 400 μm diameter, respectively. (b) Dependence of electron temperature on off focus distance. Along laser propagation axis, the point above nozzle center is defined as zero.

III to $T_e = 4.9$ MeV and ~ 70 MeV cutoff energy for Case I. Note that the shock nozzles produce symmetrical triangle-like gas distribution, the effective gas length along laser propagation axis is $2L_{grad}$. One may notice the discrepancy between gas length and acceleration length (electron energy), i.e., longer gas length but lower electron energy. Additionally, it is also observed that $T_e$ have a strong dependence on the off-focus potion of the laser pulse [FIG. 4(b)]. There exists, both for Case I and II, a certain off focus position that maximizes $T_e$, focusing the laser forward/backward will lead to a drastic decrease of $T_e$ as well as cutoff energy. Within the experimental resolution, the optimal off focus length is close to corresponding $L_{grad}$ in each case. Notice that electron can gain energy from DLA provided that the laser is well guided and a plasma channel is formed. So, we anticipate that the discrepancy is close related to laser channeling effects and will give more detailed analysis in the following part.

To gain detailed insight on the observed results, we performed a series 2D PIC simulations with the code Smilei [27]. For simplicity, we consider a triangle density profile with linear gradient of 100 μm and 200 μm up/down ramps to reproduce the experimental gas jet. An ideal trapezoidal density profile with 5 μm up/down ramps and 200 μm plateau is set for comparison. The density is uniform in transverse direction in all cases. Except that, we keep other parameters the optimal experimental ones. A gaussian laser pulse with $a_0 = 15$, τ = 35 $fs$, $\lambda_0 = 0.8$ μm, $w_0 = 3.5$ μm is considered. Neutral nitrogen is tunnel ionized to a peak plasma density of $n_e = 0.3\ n_c$ with 9 particles per cell. The moving window is 100 μm ×50 μm in laser propagation direction and transverse direction, while the corresponding resolution is set to $\lambda_0/16$ in both directions.

FIG. 5(a) shows a snapshot of electron energy density map for 100 μm ramp normalized by $n_c m_e c^2$. Majority of the energy is located in a ~ 7 μm long electron beam around the plasma channel center. Specifically, this beam presents a zig zag structure with a bunching period close to the laser wavelength. This feature distinctly suggests that DLA is the dominant acceleration regime under our experimental condition [6,7]. The highest density region with $n_e \sim 0.4 - 0.5 n_c$ appears in the outer edge of the beam [7], which is slightly higher than the background plasma density $0.3 n_c$. According to FIG. 5(a), the maximum betatron amplitude $r_\beta$ is 2 μm, which is about half of the former estimation. This can be understood by noticing that high energy electrons contribute much more to the total radiation but possess smaller $r_\beta$. Thus, adopting an average energy $\langle \gamma_e \rangle = T_e$, which means electrons with different energy are equally weighted, will lead to an overestimation of $r_\beta$. Then we can deduce the corresponding betatron strength parameter, for case I and II in FIG. 4(a), $K_\beta = \gamma_e \omega_\beta r_\beta / c$ increases from 13.6 to 22.5, which is in good agreement with the electron transverse momentum in PIC simulations. It is noted that this wiggling strength is not as strong as that reported by sub-ps PW laser [8]. One possible explanation is that 35 fs pulse duration is relatively short as stated by the self-similar theory [28,29], which limit the effective time of electron and laser resonance. Therefore, despite increasing electron energy, we can also expect more x-rays by using longer laser pulse to prolong the resonance for larger $r_\beta$ and $K_\beta$.

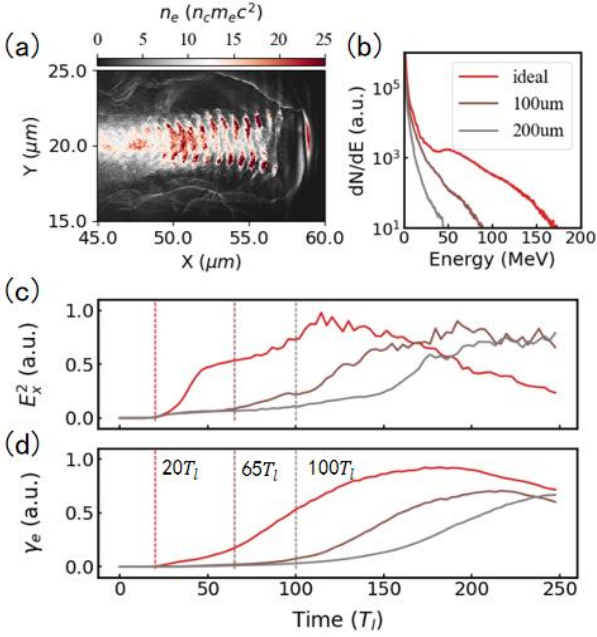

FIG. 5. Result of 2D Smilei particle in cell simulations. (a) Snapshot of electron energy density normalized to $n_c m_e c^2$ at $t = 185.6\, T_l$. (b) Electron energy spectrum for three different density ramps at $t = 123.7\, T_l$. (c), (d) Evolution of channel field energy and average electron energy.

The dependence of electron energy on gas density gradient is also reproduced in the simulations [FIG. 5(b)]. The ideal case with 5 μm up/down ramps generates the most energetic electron beam while the electron temperature and cutoff energy decrease with the ramp length, consistent with the experimental measurements. In order to illustrate the effects of density gradient in detail, the evolution of channel field energy $E_x^2$ and average electron energy $\gamma_e$ is plotted in FIG. 5(c), (d). The acceleration of electron is directly related to the formation of plasma channel, which is the basis of DLA [6]. The unique feature here is that the time when effective channel is formed differs according to the density gradient, as depicted by the vertical dotted lines, $t = 20, 65, 100\, T_l$ for the ideal and 100 μm, 200 μm ramps, respectively. Shorter density ramp enables earlier formation of plasma channel thus may lead to a longer acceleration length and eventually, higher electron energy. Plasma channel formation turned out be a consequence of several competitive effects [30], e.g., diffraction, self-focusing and ionization refraction [31]. And ionization-induced refraction has been identified an important factor in the PIC simulations, which has usually been omitted in simulation studies [6,7]. Simulations with fully ionized plasma generated opposite results that 200 μm ramps case produced most energetic beam. The dependence of electron energy with off-focus position is then understood as the balanced compensation of intensity-dependent self-focusing to ionization refraction at the optimal position. Focusing the laser forward or backward will both lower the intensity thus break the balance.

In conclusion, we have demonstrated high conversion efficiency betatron radiation from fs petawatt laser irradiated near critical density plasma. X-ray and electron diagnoses are found to be in good agreement and also with 2D particle in cell simulations. The measured x-ray energy conversion efficiency is 1-3 orders of magnitudes enhanced over other regimes, which is attributed to the high electron charge and high betatron oscillation strength by direct laser acceleration. The effects of density gradient have also been identified and interpreted with PIC simulations. Ultrahigh flux, micro-sized femtosecond x-ray source from this regime may facilitate single shot measurement of ultrafast x-ray absorption spectroscopy [2] and μCT [3,4] applications.

This work was supported by the Key Program of CAS (XDB16010200, XDA01020304, XDB17030500), the Science Challenge Project (TZ2018005), the National Key R&D Program of China (2017YFA0403301), the National Natural Science Foundation of China (11805266, 11991073, 11721404, 11905289, 61975229).

[1] S. Corde, K. Ta Phuoc, G. Lambert, R. Fitour, V. Malka, A. Rousse, A. Beck, and E. Lefebvre, Rev. Mod. Phys. **85**, 1 (2013).
[2] B. Mahieu, N. Jourdain, K. T. Phuoc, F. Dorchies, J. P. Goddet, A. Lifschitz, P. Renaudin, and L. Lecherbourg, Nature Communications **9** (2018).
[3] A. Döpp, L. Hehn, J. Götzfried, J. Wenz, M. Gilljohann, H. Ding, S. Schindler, F. Pfeiffer, and S. Karsch, Optica **5** (2018).
[4] J. M. Cole, D. R. Symes, N. C. Lopes, J. C. Wood, K. Poder, S. Alatabi, S. W. Botchway, P. S. Foster, S. Gratton, S. Johnson, C. Kamperidis, O. Kononenko, M. De Lazzari, C. A. J. Palmer, D. Rusby, J. Sanderson, M. Sandholzer, G. Sarri, Z. Szoke-Kovacs, L. Teboul, J. M. Thompson, J. R. Warwick, H. Westerberg, M. A. Hill, D. P. Norris, S. P. D. Mangles, and Z. Najmudin, Proc Natl Acad Sci U S A **115**, 6335 (2018).
[5] E. Esarey, B. A. Shadwick, P. Catravas, and W. P. Leemans, Phys Rev E Stat Nonlin Soft Matter Phys **65**, 056505 (2002).
[6] A. Pukhov, Z. M. Sheng, and J. Meyer-ter-Vehn, Physics of Plasmas **6**, 2847 (1999).
[7] B. Liu, H. Y. Wang, J. Liu, L. B. Fu, Y. J. Xu, X. Q. Yan, and X. T. He, Phys. Rev. Lett. **110**, 045002 (2013).
[8] S. Kneip, S. R. Nagel, C. Bellei, N. Bourgeois, A. E. Dangor, A. Gopal, R. Heathcote, S. P. Mangles, J. R. Marques, A. Maksimchuk, P. M. Nilson, K. T. Phuoc, S. Reed, M. Tzoufras, F. S. Tsung, L. Willingale, W. B. Mori, A. Rousse, K. Krushelnick, and Z. Najmudin, Phys. Rev. Lett. **100**,

105006 (2008).

[9] H. Y. Wang, B. Liu, X. Q. Yan, and M. Zepf, Physics of Plasmas **22** (2015).

[10] K. N. Zhou, X. J. Huang, X. M. Zeng, N. Xie, Y. L. Zuo, Y. Guo, D. B. Jiang, W. J. Dai, Q. Xue, W. Q. Huang, E. D. Li, L. Wei, L. Sun, X. Wang, and J. Q. Su, Laser Physics **28** (2018).

[11] L. Rovige, J. Huijts, A. Vernier, I. Andriyash, F. Sylla, V. Tomkus, V. Girdauskas, G. Raciukaitis, J. Dudutis, V. Stankevic, P. Gecys, and J. Faure, Rev. Sci. Instrum. **92** (2021).

[12] I. V. Khutoretsky, Rev. Sci. Instrum. **66**, 773 (1995).

[13] F. Albert, N. Lemos, J. L. Shaw, B. B. Pollock, C. Goyon, W. Schumaker, A. M. Saunders, K. A. Marsh, A. Pak, J. E. Ralph, J. L. Martins, L. D. Amorim, R. W. Falcone, S. H. Glenzer, J. D. Moody, and C. Joshi, Phys. Rev. Lett. **118**, 134801 (2017).

[14] A. L. Meadowcroft, C. D. Bentley, and E. N. Stott, Rev. Sci. Instrum. **79** (2008).

[15] S. Kneip, C. McGuffey, J. L. Martins, S. F. Martins, C. Bellei, V. Chvykov, F. Dollar, R. Fonseca, C. Huntington, G. Kalintchenko, A. Maksimchuk, S. P. D. Mangles, T. Matsuoka, S. R. Nagel, C. A. J. Palmer, J. Schreiber, K. T. Phuoc, A. G. R. Thomas, V. Yanovsky, L. O. Silva, K. Krushelnick, and Z. Najmudin, Nature Physics **6**, 980 (2010).

[16] L. M. Chen, W. C. Yan, D. Z. Li, Z. D. Hu, L. Zhang, W. M. Wang, N. Hafz, J. Y. Mao, K. Huang, Y. Ma, J. R. Zhao, J. L. Ma, Y. T. Li, X. Lu, Z. M. Sheng, Z. Y. Wei, J. Gao, and J. Zhang, Sci Rep **3**, 1912 (2013).

[17] K. Huang, Y. F. Li, D. Z. Li, L. M. Chen, M. Z. Tao, Y. Ma, J. R. Zhao, M. H. Li, M. Chen, M. Mirzaie, N. Hafz, T. Sokollik, Z. M. Sheng, and J. Zhang, Sci Rep **6**, 27633 (2016).

[18] S. Cipiccia, M. R. Islam, B. Ersfeld, R. P. Shanks, E. Brunetti, G. Vieux, X. Yang, R. C. Issac, S. M. Wiggins, G. H. Welsh, M.-P. Anania, D. Maneuski, R. Montgomery, G. Smith, M. Hoek, D. J. Hamilton, N. R. C. Lemos, D. Symes, P. P. Rajeev, V. O. Shea, J. M. Dias, and D. A. Jaroszynski, Nature Physics **7**, 867 (2011).

[19] C. H. Yu, J. S. Liu, W. T. Wang, W. T. Li, R. Qi, Z. J. Zhang, Z. Y. Qin, J. Q. Liu, M. Fang, K. Feng, Y. Wu, L. T. Ke, Y. Chen, C. Wang, Y. Xu, Y. X. Leng, C. Q. Xia, R. X. Li, and Z. Z. Xu, Appl. Phys. Lett. **112** (2018).

[20] J. M. Cole, J. C. Wood, N. C. Lopes, K. Poder, R. L. Abel, S. Alatabi, J. S. Bryant, A. Jin, S. Kneip, K. Mecseki, D. R. Symes, S. P. Mangles, and Z. Najmudin, Sci Rep **5**, 13244 (2015).

[21] J. C. Wood, D. J. Chapman, K. Poder, N. C. Lopes, M. E. Rutherford, T. G. Whites, F. Albert, K. T. Behm, N. Booth, J. S. J. Bryant, P. S. Foster, S. Glanzer, E. l. Hill, K. Krushelnick, Z. Najmudin, B. B. Pollock, S. Rose, W. Schumaker, R. H. H. Scott, M. Sherlock, A. G. R. Thomas, Z. Zhao, D. E. Eakins, and S. P. D. Mangles, Scientific Reports **8** (2018).

[22] M. Schnell, A. Savert, I. Uschmann, M. Reuter, M. Nicolai, T. Kampfer, B. Landgraf, O. Jackel, O. Jansen, A. Pukhov, M. C. Kaluza, and C. Spielmann, Nature Communications **4** (2013).

[23] W. Yan, L. Chen, D. Li, L. Zhang, N. A. Hafz, J. Dunn, Y. Ma, K. Huang, L. Su, M. Chen, Z. Sheng, and J. Zhang, Proc Natl Acad Sci U S A **111**, 5825 (2014).

[24] Y. F. Li, J. Feng, J. H. Tan, J. G. Wang, D. Z. Li, K. G. Dong, X. H. Zhang, B. Zhu, F. Tan, Y. C. Wu, Y. Q. Gu, and L. M. Chen, High Energy Density Physics **37** (2020).

[25] C. Gahn, G. D. Tsakiris, A. Pukhov, J. Meyer-ter-Vehn, G. Pretzler, P. Thirolf, D. Habs, and K. J. Witte, Phys. Rev. Lett. **83**, 4772 (1999).

[26] A. Pukhov and J. MeyerterVehn, Phys. Rev. Lett. **76**, 3975 (1996).

[27] J. Derouillat, A. Beck, F. Perez, T. Vinci, M. Chiaramello, A. Grassi, M. Fle, G. Bouchard, I. Plotnikov, N. Aunai, J. Dargent, C. Riconda, and M. Grech, Computer Physics Communications **222**, 351 (2018).

[28] S. Gordienko and A. Pukhov, Physics of Plasmas **12**, 043109 (2005).

[29] T. W. Huang, C. T. Zhou, A. P. L. Robinson, B. Qiao, A. V. Arefiev, P. A. Norreys, X. T. He, and S. C. Ruan, Physics of Plasmas **24** (2017).

[30] E. Esarey, P. Sprangle, J. Krall, and A. Ting, IEEE J. Quantum Electron. **33**, 1879 (1997).

[31] A. J. Mackinnon, M. Borghesi, A. Iwase, M. W. Jones, G. J. Pert, S. Rae, K. Burnett, and O. Willi, Phys. Rev. Lett. **76**, 1473 (1996).